\documentclass{iaus}
\usepackage{graphicx}

\def \ga{\mathrel{\mathchoice   {\vcenter{\offinterlineskip\halign{\hfil
$\displaystyle##$\hfil\cr>\cr\sim\cr}}}
{\vcenter{\offinterlineskip\halign{\hfil$\textstyle##$\hfil\cr
>\cr\sim\cr}}}
{\vcenter{\offinterlineskip\halign{\hfil$\scriptstyle##$\hfil\cr
>\cr\sim\cr}}}
{\vcenter{\offinterlineskip\halign{\hfil$\scriptscriptstyle##$\hfil\cr
>\cr\sim\cr}}}}}

\title[Cosmology and the Hubble Constant] 
{Cosmology and the Hubble Constant: \\ On the Megamaser Cosmology Project (MCP)}

\author[C. Henkel, J.A. Braatz, M. Reid et al.]   
{C. Henkel$^{1}$, J.A. Braatz$^2$, M.J. Reid$^3$, J.J. Condon$^{4}$, K.Y. Lo$^{5}$,
C.M.V. Impellizzeri$^{6}$ \and C.Y. Kuo$^7$}

\affiliation{$^1$Max-Planck-Institut f{\"u}r Radioastronomie, Auf dem H{\"u}gel 69,
             53121 Bonn, Germany \\ 
             Astron. Dept., King Abdulaziz University, P.O. Box 80203,
             Jeddah, Saudi Arabia \\ 
             email: {\tt chenkel@mpifr-bonn.mpg.de} \\[\affilskip]
             $^2$National Radio Astronomy Observatory, 520 Edgemont Road, \\
             Charlottesville, VA 22903, USA \\
             email: {\tt jbraatz@nrao.edu} \\[\affilskip]
             $^3$Harvard-Smithonsian Center for Astrophysics, 60 Garden Street, \\
             Cambridge, MA02138, USA \\
             email: {\tt reid@cfa.harvard.edu} \\[\affilskip]
             $^4$National Radio Astronomy Observatory, 520 Edgemont Road, \\
             Charlottesville, VA 22903, USA \\
             email: {\tt jcondon@nrao.edu} \\[\affilskip]
             $^5$National Radio Astronomy Observatory, 520 Edgemont Road, \\
             Charlottesville, VA 22903, USA \\
             email: {\tt flo@nrao.edu} \\[\affilskip]
             $^6$National Radio Astronomy Observatory, 520 Edgemont Road, 
             Charlottesville, \\
             VA 22903, USA \\
             Joint ALMA Observatory, Alonso de C{\'o}rdova 3107, Vitacura, 
             Santiago, Chile \\
             email: {\tt violette@nrao.edu} \\[\affilskip]
             $^7$Dept. of Astronomy, University of Virginia, Charlottesville, 
             VA 22904, USA \\
             AASIA, Astron.-Math. Building, Roosevelt Rd, Taipei 10617, Taiwan \\
             email: {\tt ck2v@virginia.edu} \\[\affilskip]
}

\pubyear{2012} \volume{287}  
\pagerange{001--012} 
 \jname{Cosmic Masers - from OH to H$_0$}
\editors{R.S. Booth, E.M.L. Humphreys \& W.H.T. Vlemmings, eds.}

\begin{document}

\maketitle

\begin{abstract}
The Hubble constant $H_0$ describes not only the expansion of local space
at redshift $z$ $\sim$ 0, but is also a fundamental parameter determining
the evolution of the universe. Recent measurements of $H_0$ anchored on 
Cepheid observations have reached a precision of several percent. However, this 
problem is so important that confirmation from several methods is needed to 
better constrain $H_0$ and, with it, dark energy and the curvature of space. 
A particularly {\it direct} method involves the determination of distances 
to local galaxies far enough to be part of the Hubble flow through water 
vapor (H$_2$O) masers orbiting nuclear supermassive black holes. The goal 
of this article is to describe the relevance of $H_0$ with respect to 
fundamental cosmological questions and to summarize recent progress of 
the ``Megamaser Cosmology Project'' (MCP) related to the Hubble constant. 

\end{abstract}

\firstsection 
\section{Cosmological Background}

For 85 years, it has been known that our universe is expanding (Lema{\^i}tre 1927).
This expansion was believed to slow down in time because of gravitational attraction. 
However, based on observations of luminous standard candles (Type Ia supernoave) 
Riess et al. (1998) and Perlmutter et al. (1999) suggested instead accelerated 
expansion, turning cosmology upside down and winning the most recent Nobel Prize in 
physics. More than a decade after this discovery, accelerated expansion is well 
established. A de-acceleration during the initial few billion years after the 
``Big Bang'', when densities of matter and radiation were much higher than today, 
is followed by accelerated expansion. The cause of the accelerated expansion is 
so far unknown and is described by the term ``Dark Energy''. Following the standard 
model, it should account for the majority of the energy density of the universe and 
retards the formation of large scale structure. Understanding dark energy may be 
the most important problem existing in physics today. 

Dark Energy dominates the energy budget, accelerates the expansion of the universe, and 
affects large scale structure. What is its nature? There are three classes of 
potential explanations: (1) a cosmological constant, which has been proposed already 
in the early days of general relativity (Einstein 1917) as a kind of repulsive
gravity, (2) a scalar field, somewhat analogous to that proposed to explain inflation at
a much earlier time (e.g., Wetterich 1988; Ratra \& Peebles 1988), and (3) modified
gravity (e.g., Tsujikawa 2010), which will not be considered here. 

Assuming that the universe is homogeneous and isotropic (as approximately suggested 
by the large scale matter distribution and the 3\,K microwave background), the 
space-time metric can be written in the following form
\begin{equation}
{\rm d}s^2 = {\rm d}t^2 - 
a^2(t)\,\times\,[{\rm d}r^2/(1-kr^2) + r\,{\rm d}\,\theta^2 + 
r^2\,{\rm sin}^2\theta\,\,{\rm d}\phi^2], 
\end{equation}
with $t$ and $a$ $\propto$ (1 + $z$)$^{-1}$ being time and cosmic scale factor, $r$, 
$\theta$, and $\phi$ denoting comoving spatial coordinates, and $k$ representing 
the curvature of 3-dimensional space. The field equations of general relativity, 
applied to this Friedmann-Robertson-Walker metric, lead to the so-called Friedmann 
equations,
\begin{equation}
H = \left(\frac{\dot{a}}{a}\right)^2 = \frac{8\,\pi\,{\rm G}}{3{\rm c}^2}\,\rho - 
k\,\frac{{\rm c}^2}{a^2} + \frac{\Lambda}{3}
\end{equation}
and
\begin{equation}
\frac{\ddot{a}}{a} = -\frac{4\,\pi\,{\rm G}}{3{\rm c}^2}\,(\rho + 3p) + \frac{\Lambda}{3}.
\end{equation}
$H$ is the Hubble parameter for a given time ($H_0$ stands for redshift $z$ = 0), $\rho$ is 
the density of matter and radiation, $p$ denotes pressure, and $\Lambda$ represents the 
traditional cosmological constant, which (like dark energy) can be subsumed into the 
density and pressure term,
\begin{equation}
\frac{\ddot{a}}{a} = -\frac{4\,\pi\,{\rm G}}{3{\rm c}^2}\,\,\,\sum\limits_{i}^{n}(\rho_{\rm i} + 3p_{\rm i}).
\end{equation}
Defining $w$ = $p$/$\rho$, the different (in part putative) components yield: \\
\begin{center}
Matter: $w$ = 0 \\ Radiation: $w$ = 1/3 \\ ``Quintessential'' scalar field: --1 $<$ 
$w$ $<$ --1/3 \\ Cosmological constant: $w$ = --1 \\ Phantom energy: $w$ $<$ --1 \\
\end{center}

\bigskip
The resulting values of $w$ determine the normalized acceleration $\ddot{a}$/$a$. For 
gravity (matter) we obtain the expected negative value. This also holds for radiation, which 
has dominated, according to the standard model, at redshifts $\ga$10$^4$. The ($\rho$ +
3$p$) term in Eq.\,1.4 directly infers that $w$ values smaller than --1/3 are required 
for accelerated expansion, which is therefore the possible $w$ range for dark energy. $w$ 
and $H$, and thus also $H_0$, are obviously related, emphasizing the cosmological importance 
of the Hubble constant.

\begin{figure}[t]
\hspace{0.0 cm}
\begin{center}
\includegraphics[width=12.7cm]{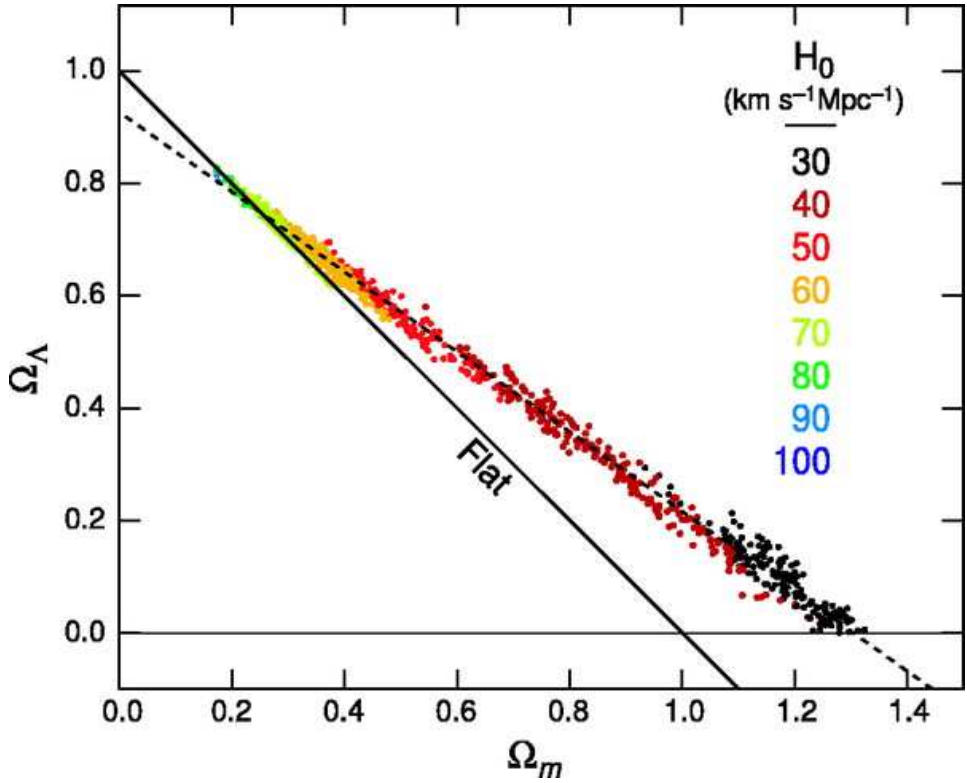} 
\hspace{0.0 cm}
\caption{Constraints from the cosmic microwave background (Spergel et al. 2007).
The different colors correspond to different values of the Hubble constant $H_0$.
$\Omega_{\rm m}$ is the energy density of baryonic and dark matter in units of the 
critical density, $\Omega_\Lambda$ is the corresponding parameter for dark energy. 
For a ``flat'' universe, $\Omega_{\rm m}$ + $\Omega_{\Lambda}$ = 1. The data are 
consistent with a wide range of $H_0$ values.}
\label{fig1}
\end{center}
\end{figure}

Fig.~1 shows the range of not necessarily flat cosmological cold dark matter (CDM)
models consistent with Wilkinson Microwave Anisotropy Probe (WMAP) data (Spergel et 
al. 2007). Assuming that the universe is flat would provide a rather accurate solution, 
but is it really flat as suggested by inflationary models? The present value of the 
curvature radius, $R_0$, is related to $H_0$ and $\Omega_0$ = $\Omega_{\rm m}$ + 
$\Omega_\Lambda$ (see Fig.~1 for definitions) by 
\begin{equation}
R_{0}\ \ = \ \ (a/2) \times\ |k|^{-1/2}\ \ =\ \ ({\rm c}/2) \times\ H_0^{-1} \times\ 
(\Omega_0 - 1)^{-1/2}.
\end{equation}
c is the speed of light and (c/2) $\times$ $H_0^{-1}$ $\sim$ 2.1\,Gpc is 
known as the Hubble radius. Obviously, with our present precision to determine 
$\Omega_0$, curvature radii as small 10\,Gpc cannot be excluded and we are still 
far from being able to state that the universe is truly flat. All this implies 
that the cosmic microwave background alone does not provide stringent limits. 
This is not unexpected, since the CMB provides information at a single early 
epoch, when dark energy did not play a role. Its importance can only be deduced 
by a combination of CMB observations with data from the much younger universe.

There are several lines of observational activity to constrain the wide range of 
models permitted by the CMB: These include (1) Type Ia supernovae, the standard 
candles, where luminosity distances and redshifts can be compared; (2) galaxy 
clusters, where redshift independent ratios between baryonic and dark matter 
masses can only be obtained with a small subset of possible cosmologies; (3) 
gravitational lensing or cosmic shear, where the dark matter distribution can be 
determined, isolating dark energy; and (4) baryon acoustic oscillations (BAO), 
which left an imprint on the cosmic microwave background, which is clearly seen in 
CMB power spectra (e.g., Spergel et al. 2003, 2007; Komatsu et al. 2011). This imprint 
is still visible affecting structure at moderate redshifts and providing a local 
size scale of order 140\,Mpc. All these tracers are observed at significant 
redshifts. However, it is at redshift zero where dark energy is most dominant, 
and only here its energy density is significantly higher than that of baryonic and 
dark matter combined. Thus it is the Hubble constant, providing a measure of 
the {\it local} universe, which provides the longest lever arm with respect to 
the CMB to measure the effects of dark energy (Hu 2005).

\begin{figure}[t]
\hspace{0.0 cm}
\begin{center}
\includegraphics[width=14.3cm]{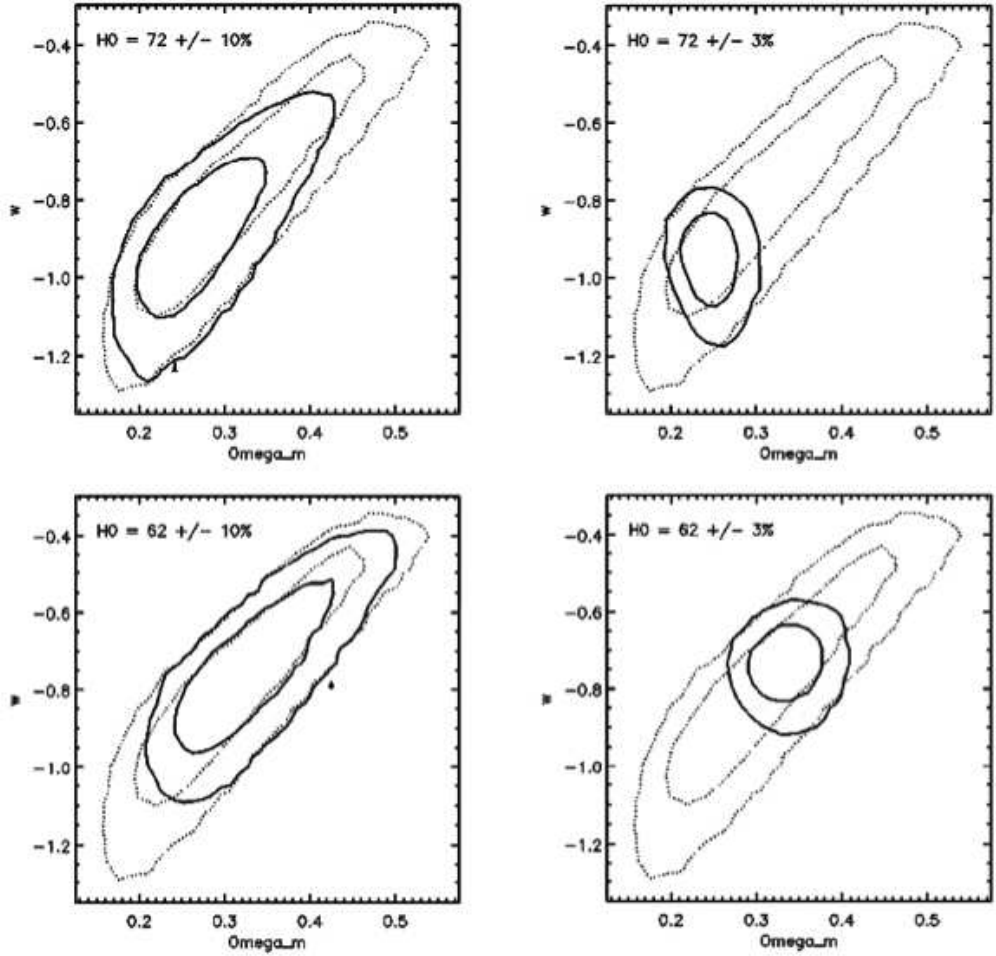} 
\hspace{0.0 cm}
\caption{WMAP (Wilkinson Microwave Anisotropy Probe) 1$\sigma$ and 2$\sigma$ 
likelihood surfaces for $\Omega_{\rm m}$ (see Fig.~1) and $w$ (Eq.\,1.4), 
with priors on $H_0$ (upper panels: 72\,km\,s$^{-1}$\,Mpc$^{-1}$; lower panels: 
62\,km\,s$^{-1}$\,Mpc). Right versus left panels (solid lines) demonstrate the 
improvements gained by reducing the uncertainty in $H_0$ from 10\% to 3\%. Dotted 
lines: ``wcdm + no perturbations'' model from Spergel et al. (2007); solid lines: 
the same, but with constraints from $H_0$ incorporated.}
\label{fig2}
\end{center}
\end{figure}

\section{Constraining $H_0$}

Assuming a $\Lambda$ cold dark matter universe and excluding curvature, 
exotic neutrino or specific early universe physics, Komatsu et al. (2011) 
derive from WMAP data $H_{\rm 0}$ = 71.0 $\pm$ 2.5\,km\,s$^{-1}$\,Mpc$^{-1}$. 
While studies based on gravitational lens time delays (e.g., Treu \& 
Koopmans 2002; Cardone et al. 2002) and on X-ray and Sunyaev-Zel'dovich data
of galaxy clusters (e.g., Bonamente et al. 2006) have been used to constrain $H_0$, 
such deductions from redshifted objects depend on the chosen cosmological 
model and are no substitute for a measurement of $H_0$ in the local 
universe. 

With the HST key project to measure the Hubble constant, Freedman et al. 
(2001) obtained from Cepheids in nearby galaxies $H_0$ = 72 $\pm$ 3$_{\rm r}$ 
$\pm$ 7$_{\rm s}$\,km\,s$^{-1}$\,Mpc$^{-1}$, estimating both random and 
systematic errors. This was based on the extragalactic distance ladder 
using Cepheid variable stars in the Large Magellanic Cloud to calibrate 
the measurements. However, the LMC has a low metallicity and significant 
depth. A potential dependence of Cepheid luminosities on metallicity and 
uncertainties in the distances to individual stars complicate the calibration. 
This was highlighted by Sandage et al. (2006), who used similar methods 
but a different metallicity correction to obtain $H_0$ = 62 $\pm$ 
1.3$_{\rm r}$ $\pm$ 5.0$_{\rm s}$ km\,s$^{-1}$\,Mpc$^{-1}$ (Fig.~2). While 
the $H_0$ value by Freedman et al. (2001) is consistent with a flat universe, 
the Sandage et al. (2006) result challenges it. 

To date the most ambitious program to determine $H_0$ is that led by A.G. Riess
(see also Freedman \& Madore 2010 for a recent review). Based on three anchors, 
(1) the parallax determinations of Galactic Cepheids, (2) Cepheides in the 
LMC with distances deduced from eclipsing binaries, (3) the distance to NGC\,4258 
(see Sect.\,3) and with new {\it HST} (Hubble Space Telescope) data from galaxies 
with Cepheids {\it and} Type Ia supernovae, Riess et al. (2011) derive $H_0$ = 
73.8 $\pm$ 2.4\,km\,s$^{-1}$\,Mpc$^{-1}$. This estimate also makes use of
Cepheids measured in the near infrared, which helps to reduce both systematic and 
random errors with respect to optical observations. 
Note that the distance to 
NGC\,4258 has been slightly revised in the meantime (E.M.L. Humphreys, priv. 
comm.). 

While all these measurements are highly encouraging and pave the way to a more 
precise knowledge of our universe, a totally independent measure of $H_0$ is 
essential to either confirm the above cited results or to hint at problems
that may have been overlooked so far.

\section{The Megamaser Cosmology Project (MCP)}

The MCP is an NRAO (National Radio Astronomy Observatory) key project to
determine $H_0$ by measuring geometric distances with an accuracy of 
$\sim$10\% to $\sim$10 galaxies in the local Hubble flow. Following the 
experience gained by studying the prototypical source, NGC~4258 (Miyoshi
et al. 1995; Herrnstein et al. 1999), it includes (1) a GBT (Green 
Bank Telescope) survey to identify suitable circumnuclear 22\,GHz H$_2$O 
maser disks (see Fig.~3 for a spectrum), (2) direct imaging of these sub-pc disks, using 
the {\it VLBA} (Very Long Baseline Array), the {\it GBT}, and, for northern 
sources, also the {\it Effelsberg} telescope, (3) {\it GBT} monitoring to 
measure accelerations of the spectral components, and (4) model calculations 
to simulate the maser disk dynamics. So far published articles include Reid 
et al. (2009), Braatz et al. (2010), Greene et al. (2010), and Kuo et al. 
(2011).

\begin{figure}[t]
\hspace{-1.9 cm}
\includegraphics[width=16.0cm]{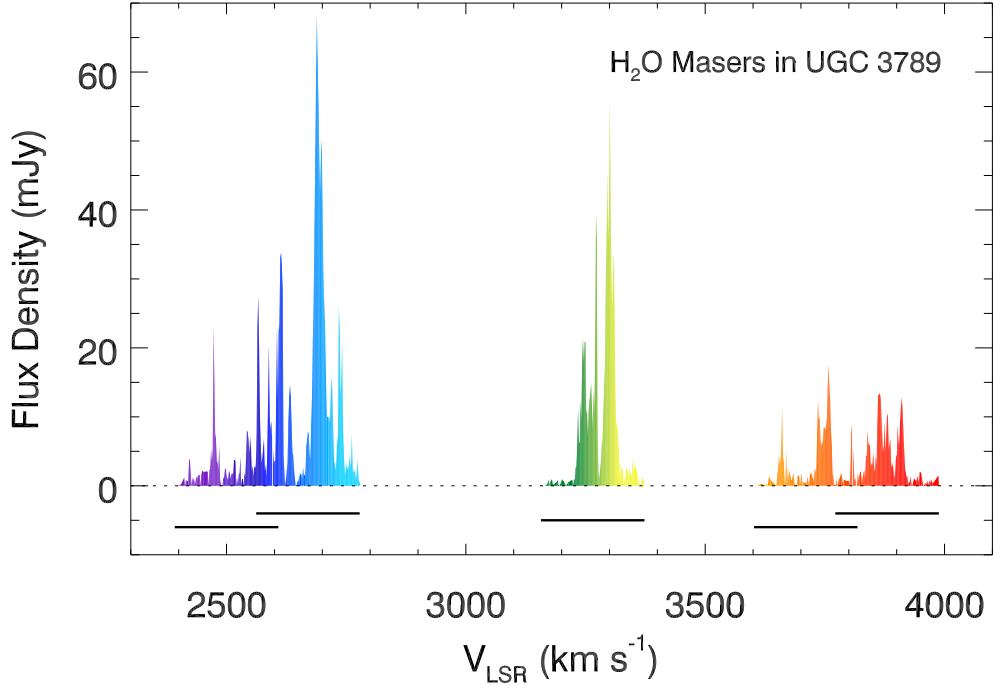} 
\vspace{-1.0 cm}
\caption{One of the newly found maser disks with systemic (green-yellow)
and high velocity (violet-blue and orange-red) components.} 
\label{fig3}
\end{figure}

The 22\,GHz H$_2$O ``megamasers'', luminous masers associated with active 
galactic nuclei, are mostly found in Seyfert 2 and LINER (Low Ionization 
Nuclear Emission Line Region) galaxies with high column densities (Braatz et 
al. 1997; Zhang et al. 2006; Madejski et al. 2006; Greenhill et al. 2008), 
relatively high optical luminosity, velocity dispersion, and [O{\sc{iii}}]$\lambda$5007
luminosity (Zhu et al. 2011) as well as relatively strong Fe K$\alpha$ 
lines in those sources which are Compton thick (Zhang et al. 2010). Low 
X-ray/[O{\sc{iv}}]$\lambda$25890 ratios (Ramolla et al. 2011), and high
nuclear radio continuum luminosities (Zhang et al. 2012) with respect 
to H$_2$O undetected galaxies are also statistically obtained. 

The H$_2$O masers reside in thin, edge-on gaseous annuli. Emission near the 
systemic velocity of the parent galaxy originates from the near side of
the disk and red- and blue-shifted satellite lines come from the two tangent 
points (see Figs.~3--5). Assuming an ideal circular, warpless thin disk, seen 
perfectly edge-on, and in Keplerian motion, the mass $M_{\rm AGN}$ enclosed 
by the disk is
\begin{equation}
M_{\rm AGN} = 1.12 \times\ \left[\frac{V_{\rm rot}}{\rm km\,s^{-1}}\right]^2 \times\
\left[\frac{R}{\rm mas}\right] \times\ \left[\frac{D}{\rm Mpc}\right]\ \,{\rm M}_{\odot},
\end{equation}
with $V_{\rm rot}$ denoting the rotation velocity at angular radius $R$ and $D$
representing the distance. Very long baseline interferometry maps allow us
to directly measure $V_{\rm rot}$ for various values of $R$. From the Keplerian
rotation curve we then obtain the constant
\begin{equation}
C_1 = \left[\frac{V_{\rm rot}}{\rm km\,s^{-1}}\right] \times\ 
\left[\frac{R}{\rm mas}\right]^{1/2}.
\end{equation}
The velocity gradient of the systemic features as a function of impact parameter
provides another constant,
\begin{equation}
C_2 = \left[\frac{V_{\rm rot}}{\rm km\,s^{-1}}\right] \times\ 
\left[\frac{R}{\rm mas}\right]^{-1}.
\end{equation}
$C_1$/$C_2$ = $R^{3/2}$ then gives the angular radius $R_{\rm s}$ of the systemic
features as viewed from a direction in the plane of the disk, but perpendicular to 
the line of sight. The total distance to the galaxy is then determined by the centripetal 
acceleration
\begin{equation}
{\rm d}V_{\rm s} / {\rm d}t = \frac{V_{\rm rot}^2}{r_{\rm s}},
\end{equation}
with the index ``s'' denoting the systemic maser components. With d$V_{\rm s}$/d$t$
being measured, the linear scale $r_{\rm s}$ can be compared with the angular scale 
$R_{\rm s}$ to provide the preliminary distance estimate.

\begin{figure}[t]
\hspace{0.0 cm}
\includegraphics[width=6.2cm]{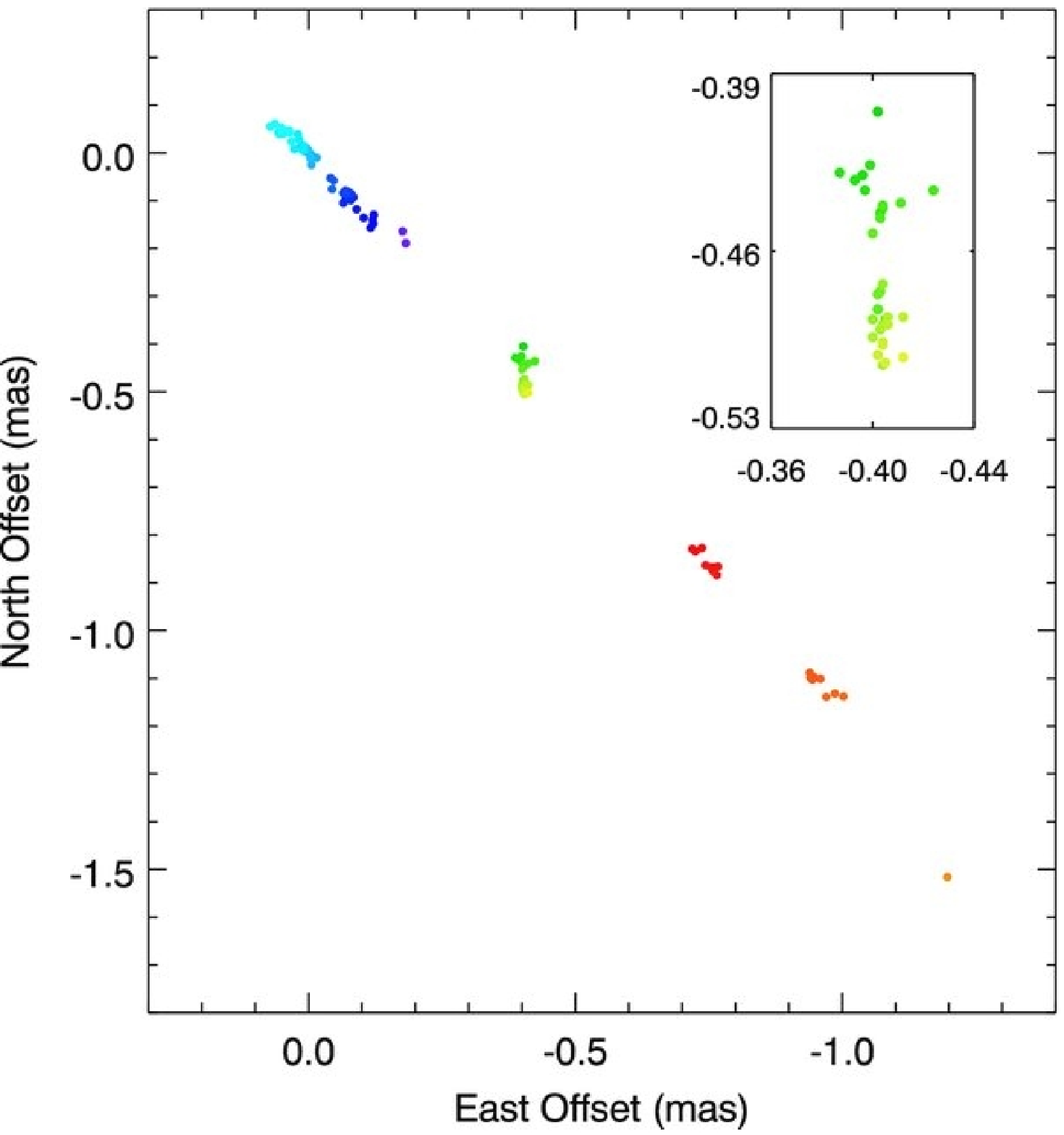} 
\vspace{0.5cm}
\includegraphics[width=7.0cm]{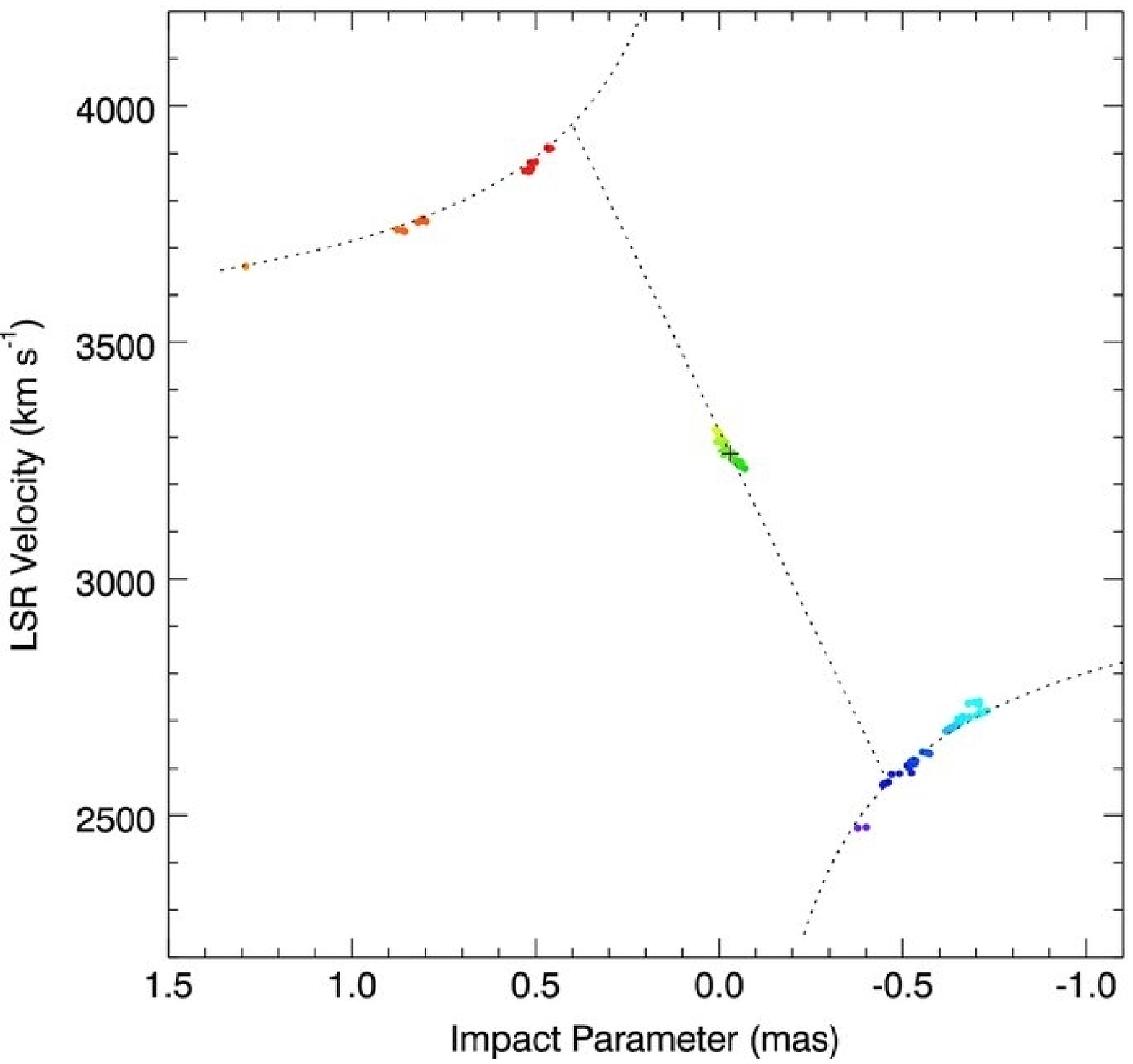} 
\vspace{0.0 cm}
\caption{Left panel: H$_2$O image of the maser disk in UGC~3789 (from Reid et 
al. 2009). The insert presents a magnification of the systemic features.
Right panel: Radial velocity versus impact parameter (also Reid 
et al. 2009). For the blue- and red-shifted high velocity components, Keplerian 
$r^{-1/2}$ rotation curves are also displayed.}
\label{fig4}
\end{figure}

While such a procedure leads to a rough first estimate, disks may be warped
(see Miyoshi et al. 1995 for the first such case, NGC~4258), orbits may 
be eccentric, and inclinations may not equal 90$^{\circ}$. To model 
the circumnuclear disks as seen in H$_2$O as detailed as possible, a 
Bayesian fitting procedure has been developed (M.J. Reid), using a 
Markov Chain Monte Carlo approach. A Metropolis Hastings algorithm 
is applied to choose successive trial parameters covering the parameter space. 
Fig.~6 displays such a simulation for NGC~6264. These very 
preliminary simulations indicate low eccentricities ($e$ $<$ 0.1).

\begin{figure}[t]
\hspace{0.0 cm}
\begin{center}
\includegraphics[width=12.6cm]{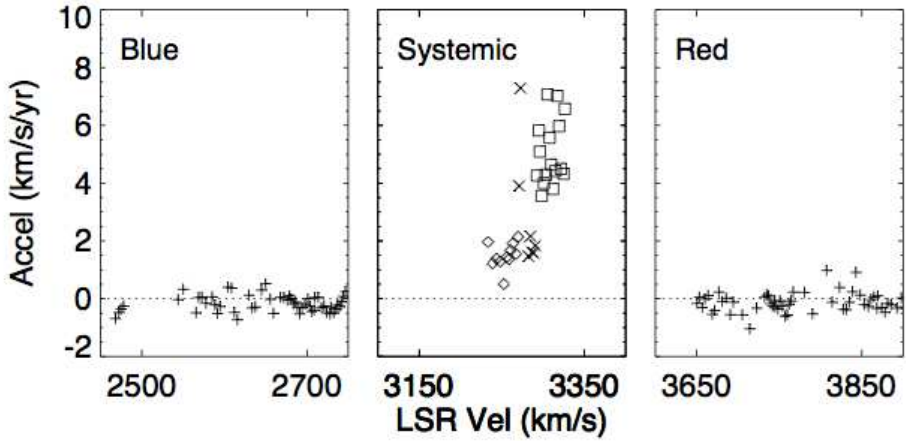} 
\caption{Acceleration of individual maser features in UGC~3789}
\label{fig5}
\end{center}
\end{figure}

For UGC~3789, Fig.~5 indicates two groups of systemic maser 
components, one with a higher acceleration than the other. Unlike 
in NGC~4258, the systemic features do not arise from a single ring segment 
with specific galactocentric radius. The distance to the galaxy can be derived 
separately for the two rings. These yield $D_1$ = 50.2 $\pm$ 7.7\,Mpc and 
$D_2$ = 48.1 $\pm$ 17.4\,Mpc. The weighted mean is $D_{\rm UGC3789}$ = 49.9
$\pm$ 7.0\,Mpc (14\%). With a peculiar radial velocity relative to the cosmic 
microwave background of --151 $\pm$ 163\,km\,s$^{-1}$ and $V_{\rm CMB}$ = 
$V_{\rm LSR}$ + 60\,km\,s$^{-1}$, the relativistic, recessional flow 
velocity becomes 3481 $\pm$ 163\,km\,s$^{-1}$. This yields with standard 
$\Lambda$CDM parameters $H_0$ = 69 $\pm$ 11\,Mpc$^{-1}$ (Braatz et al.
2010). 

For NGC\,6264, a source at a distance of 150\,Mpc, a similar 
analysis results in $H_0$ = 65.8 $\pm$ 7.2\,km\,s$^{-1}$ (Kuo 2011). 
Combining both sources, our present best estimate for the Hubble constant 
becomes $H_0$ = 67 $\pm$ 6\,km\,s$^{-1}$. This preliminary result is so 
far consistent with all previously (Sect.~2) mentioned values.

\begin{figure}[t]
\hspace{0.0 cm}
\begin{center}
\includegraphics[width=13.6cm]{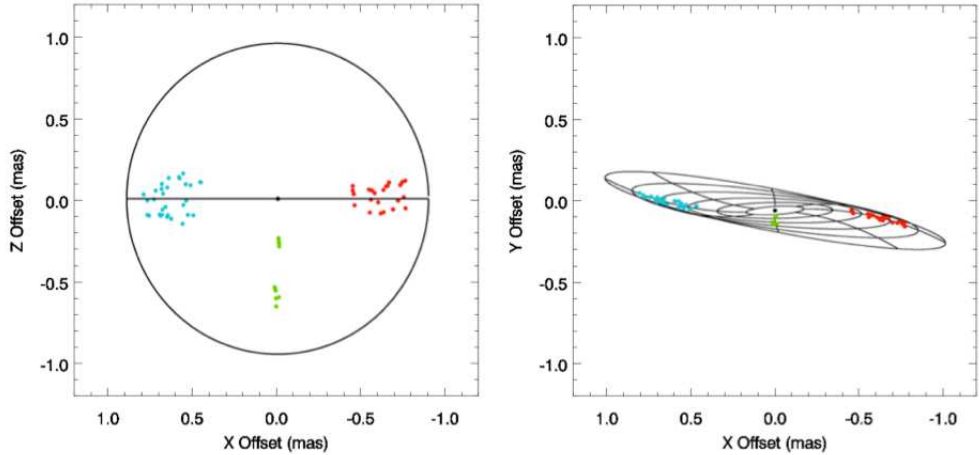} 
\caption{Bayesian fitting results of the maser disk in NGC~6264. Left panel: 
Face-on View onto the disk with systemic (green), approaching (blue)
and receding (red) components. Right panel: Model of the warped disk
viewed approximately edge-on as observed from Earth.}
\label{fig6}
\end{center}
\end{figure}

\section{Prospects}

So far, two sources yielded publishable results. Increasing
this number to 10 and accounting for the fact that the targets are located in
different parts of the sky, the 1$\sigma$ error of 6\,km\,s$^{-1}$\,Mpc$^{-1}$
obtained so far should decrease by a factor of (2/10)$^{1/2}$ to
$\sim$2.7\,km\,s$^{-1}$\,Mpc$^{-1}$ or 4\%. Longer monitoring and more interferometric
maps can reduce this uncertainty further. To demonstrate the degree of 
sensitivity required for these measurements, Fig.~7 shows the size of the 
prototypical nuclear disk in NGC~4258 on the same angular scale as the disk
toward NGC~6323. While NGC~4258 is with $V$ $\sim$ 500\,km\,s$^{-1}$ not yet 
in the Hubble flow and therefore not useful for a direct $H_0$ estimate (its 
maser lines are nevertheless essential to calibrate the distance scale defined 
by Cepheids), NGC~6323 has the potential to probe the distance scale with 
its recesseional velocity of almost 7800\,km\,s$^{-1}$. Toward NGC~6323, 
NGC~1194, NGC~2273, and Mrk~1419 the maser disks have also been mapped, 
demonstrating that the technique to derive distances, first tried out on 
NGC~4258, can also be used for much more distant galaxies.

\begin{figure}[t]
\hspace{0.0 cm}
\begin{center}
\includegraphics[width=14.6cm]{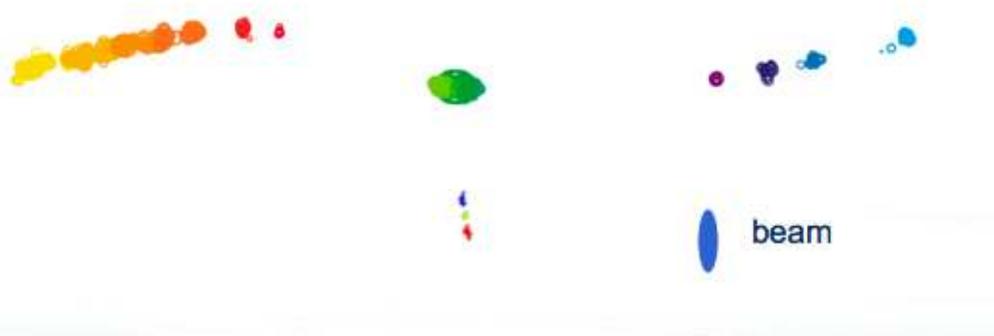} 
\caption{The maser disk of NGC~4258 (Argon et al. 2007), extended east-west, and below the 
maser disk of NGC6323, extended approximately north-south, on the same
angular scale. The given beam shows a typical synthesized beam for a 
high-declination target at 22\,GHz using the {\it VLBA}, the {\it GBT}, 
and {\it Effelsberg}. The sketch emphasizes the great progress achieved 
in recent years when mapping H$_2$O maser emission in distant sources.}
\label{fig6}
\end{center}
\end{figure}

Figs.~3, 5, and 7 directly demonstrate the importance of sensitivity.
While in UGC~3789 many features have flux densities of $\sim$5\,mJy
or higher, which can be readily analyzed, more distant sources reveal
a plethora of components below this critical level. Getting these
components as well would greatly facilitate any analysis. Thus 
the inclusion of a phased {\it Jansky VLA} (Very Large Array) 
is highly desirable. The completion of the Sardinia telescope
may also help in the foreseeable future. Furthermore, as Fig.~7 
indicates, angular resolution is another essential point. While 
an SKA-high would guarantee extreme sensitivity, the small
angular extent of the H$_2$O maser disks requires a world-wide array,
with space-VLBI providing another significant improvement.

Aside of the maser sources mentioned above, new targets have been
detected, which look promising when analyzing their single dish spectra. 
While it remains to be seen how useful they will be for detailed 
mapping, it is worth mentioning that so far no systemic feature has 
been detected that shows a secular drift to the blue side. Either 
the nuclei are opaque at 22\,GHz or the radiation is so highly beamed
that maser photons from the backside of the disks have no chance to 
reach us.


\begin{thebibliography}{}

\bibitem[Argon et al. (2007)]{argon07}
{Argon, A.L., Greenhill, L.J., Reid, M.J., et al.} 2007,
\textit{ApJ}, 659, 1040

\bibitem[Bonamente et al. (2006)]{bonamente06}
{Bonamente, M., Joy, M.K., LaRoque, S.J., et al.} 2006,
\textit{ApJ}, 647, 25

\bibitem[Braatz et al. (1997)]{braatz97}
{Braatz, J.A., Wilson, A.S., Henkel, C.} 1996,
\textit{ApJS}, 110, 321

\bibitem[Braatz et al. (2010)]{braatz10}
{Braatz, J.A., Reid, M.J., Humphreys, E.M.L. et al.} 2010,
\textit{ApJ}, 718, 657

\bibitem[Cardone et al. (2002)]{cardone02}
{Cardone, V.F., Capozziello, S., Re, V., \& Piedipalumbo, E.} 2002,
\textit{A\&A}, 382, 792

\bibitem[Einstein (1917)]{einstein17}
{Einstein, A.} 1917,
\textit{Sitzungsber. K{\"o}nigl. Preu{\ss}. Akad. der Wiss.}, 6, 142

\bibitem[Freedman \& Madore (2010)]{freedman10}
{Freedman, W.L., \& Madore, B.F.} 2010,
\textit{ARA\&A}, 48, 673

\bibitem[Freedman et al. (2001)]{freedman01}
{Freedman, W.L., Madore, B.F., Gibson, B.K., et al.} 2001,
\textit{ApJ}, 553, 47

\bibitem[Greenhill et al. (2008)]{greenhill08}
{Greenhill, L.J., Tilak, A., \& Madejski, G.} 2008,
\textit{ApJ}, 686, L13

\bibitem[Greene et al. 2010]{greene10}
{Greene, J.E., Peng, C.Y., Kim, M. et al.} 2010,
\textit{ApJ}, 721, 26

\bibitem[Herrnstein et al. (1999)]{herrnstein99}
{Herrnstein, J.R., Moran, J.M., Greenhill, L.J. et al.} 1999,
\textit{Nature}, 400, 539

\bibitem[Hu (2005)]{hu05}
{Hu, W.} 2005,
\textit{ASP Conf. Ser.} 339, Observing Dark Energy, eds.
S.C. Wolff \& T.R.Lauer (San Francisco, ASP), 215

\bibitem[Komatsu et al. (2011)]{komatsu11}
{Komatsu, E., Smith, K.M., Dunkley, J., et al.} 2011,
\textit{ApJS}, 192, 1

\bibitem[Kuo, C.Y. 2011]{kuo11a}
{Kuo, C.Y.} 2011,
\textit{Ph.D. Thesis}, Univ. of Virginia, Charlottesville

\bibitem[Kuo, C.Y. et al. 2011]{kuo11}
{Kuo, C.Y., Braatz, J.A., Condon, J.J. et al.} 2011,
\textit{ApJ}, 727, 20

\bibitem[Lemaitre (1927)]{Lemaitre27}
{Lema{\^i}tre, G.} 1927,
\textit{Annales de la Soci{\'e}t{\'e} Scientifique de Bruxelles}, 47, 49 

\bibitem[Madejski et al. (2006)]{madejski06}
{Madejski, G., Done, C., \& Zycki, P.T.} 2006,
\textit{ApJ}, 636, 75

\bibitem[Miyoshi et al. 1995]{miyoshi95}
{Miyoshi, M., Moran, J., Herrnstein, J. et al.} 1995,
\textit{Nature}, 373, 127

\bibitem[Perlmutter et al. (2003)]{perl99}
{Perlmutter, S., Aldering, G., Goldhaber, G., et al.} 1999, 
\textit{ApJ}, 517, 565

\bibitem[Ramolla et al. (2011)]{ramolla11}
{Ramolla, M., Haas, M., Bennert, V.N., \& Chini, R.}, 2011
\textit{A\&A}, 530, 147

\bibitem[Ratra \& Peebles (1988)]{ratra88}
{Ratra, B. \& Peebles, P.J.E.} 1988
\textit{Phys. Rev. D}, 37, 3406 

\bibitem[Reid et al. (2009)]{reid09}
{Reid, M.J., Braatz, J.A., Condon, J.J, et al.} 2009,
\textit{ApJ}, 695, 287

\bibitem[Riess et al. (1998)]{riess98}
{Riess, A.G., Filippenko, A.V., Challis, P., et al.} 1998, 
\textit{AJ}, 116, 1009

\bibitem[Riess et al. (2011)]{riess11}
{Riess, A.G., Macri, L., Casertano, S., et al.} 2011, 
\textit{ApJ}, 730, 119

\bibitem[Sandage et al. 2006]{sandage06}
{Sandage, A., Tammann, G.A., Saha, A., et al.} 2006,
\textit{ApJ}, 653, 843

\bibitem[Spergel (2003)]{spergel03}
{Spergel, D.N., Verde, L., Peiris, H.V., et al.} 2003,
\textit{ApJS}, 148. 175

\bibitem[Spergel (2007)]{spergel07}
{Spergel, D.N., Bean, R., Dor{\'e}, O., et al.} 2007,
\textit{ApJS}, 170. 377

\bibitem[Treu \& Koopmans (2002)]{treu02}
{Treu, T., \& Koopmans, L.V.E.} 2002,
\textit{MNRAS}, 337, L6

\bibitem[Tsujikawa, S. (2010)]{tsujikawa11}
{Tsujikawa, S.} 2010,
\textit{Lect. Notes in Phys.}, 800, 99

\bibitem[Wetterich (1988)]{wetterich88}
{Wetterich, C.} 1988
\textit{Nucl. Phys. B}, 302, 668 

\bibitem[Zhang et al. (2006)]{zhang06}
{Zhang, J.S., Henkel, C., Kadler, M., et al.} 2006
\textit{A\&A}, 450, 933

\bibitem[Zhang et al. (2010)]{zhang10}
{Zhang, J.S., Henkel, C., Gui, Q., et al.} 2010
\textit{ApJ}, 708, 1582

\bibitem[Zhang et al. (2012)]{zhang12}
{Zhang, J.S., Henkel, C., Gui, Q., \& Wang, J.} 2012
\textit{A\&A}, 538, 152

\bibitem[Zhu et al. (2011)]{zhu11}
{Zhu, G., Zaw, I., Blanton, M.R., \& Greenhill, L.J.} 2011
\textit{ApJ}, 742, 73


\end{thebibliography}
\end{document}